\documentclass[12pt,preprint]{aastex}

\usepackage{color}

\slugcomment{To appear in AJ}

\shorttitle{Detection and tracking of space debris with the MWA}
\shortauthors{Tingay et al.}

\begin{document}


\title{On the detection and tracking of space debris using the Murchison Widefield Array. I. Simulations and test observations demonstrate feasibility}


\author{S.~J.~Tingay}
\affil{International Centre for Radio Astronomy Research, Curtin University, Perth, Australia}
\affil{ARC Centre of Excellence for All-sky Astrophysics (CAASTRO)}
\email{s.tingay@curtin.edu.au}

\author{D.~L.~Kaplan}
\affil{University of Wisconsin--Milwaukee, Milwaukee, USA}

\author{B.~McKinley, F.~Briggs}
\affil{The Australian National University, Canberra, Australia}
\affil{ARC Centre of Excellence for All-sky Astrophysics (CAASTRO)}

\author{R.B.~Wayth, N.~Hurley-Walker, J.~Kennewell}
\affil{International Centre for Radio Astronomy Research, Curtin University, Perth, Australia}

\author{C.~Smith}
\affil{Electro Optic Systems Pty Ltd, Canberra, Australia}

\author{K.~Zhang}
\affil{RMIT University, Melbourne, Australia}

\author{W.~Arcus, R.~Bhat, D.~Emrich, D.~Herne, N.~Kudryavtseva, M.~Lynch, S.M.~Ord, M.~Waterson}
\affil{International Centre for Radio Astronomy Research, Curtin University, Perth, Australia}

\author{D.G.~Barnes}
\affil{Monash e-Research Centre, Monash University, Clayton, Australia}

\author{M.~Bell, B.M.~Gaensler, E.~Lenc}
\affil{The University of Sydney, Sydney, Australia}
\affil{ARC Centre of Excellence for All-sky Astrophysics (CAASTRO)}

\author{G.~Bernardi, L.J.~Greenhill, J.C.~Kasper}
\affil{Harvard-Smithsonian Center for Astrophysics, Cambridge, USA}

\author{J.~D.~Bowman, D.~Jacobs} 
\affil{Arizona State University, Tempe, USA}

\author{J.~D.~Bunton, L.~deSouza, R.~Koenig, J.~Pathikulangara, J.~Stevens}
\affil{CSIRO Astronomy and Space Science, Australia}

\author{R.~J.~Cappallo, B.~E.~Corey, B.~B.~Kincaid, E.~Kratzenberg, C.~J.~Lonsdale, S.~R.~McWhirter, A.~E.~E.~Rogers, J.~E.~Salah, A.~R.~Whitney}
\affil{MIT Haystack Observatory, Westford, USA}

\author{A.~Deshpande, T.~Prabu, N.~Udaya~Shankar, K.~S.~Srivani, R.~Subrahmanyan}
\affil{Raman Research Institute, Bangalore, India}

\author{A.~Ewall-Wice, L.~Feng, R.~Goeke, E.~Morgan, R.~A.~Remillard, C.~L.~Williams}
\affil{MIT Kavli Institute for Astrophysics and Space Research, Cambridge, USA}

\author{B.~J.~Hazelton, M.~F.~Morales}
\affil{University of Washington, Seattle, USA}

\author{M.~Johnston-Hollitt}
\affil{Victoria University of Wellington, Wellington, New Zealand}

\author{D.~A.~Mitchell, P.~Procopio, J.~Riding, R.~L.~Webster, J.~S.~B.~Wyithe} 
\affil{The University of Melbourne, Melbourne, Australia}
\affil{ARC Centre of Excellence for All-sky Astrophysics (CAASTRO)}

\author{D.~Oberoi}
\affil{National Centre for Radio Astrophysics, Tata Institute for Fundamental Research, Pune, India}

\author{A.~Roshi} 
\affil{National Radio Astronomy Observatory, Charlottesville, USA}

\author{R.~J.~Sault} 
\affil{The University of Melbourne, Melbourne, Australia}

\author{A.~Williams} 
\affil{International Centre for Radio Astronomy Research, University of Western Australia, Perth, Australia}

\begin{abstract}
The Murchison Widefield Array (MWA) is a new low frequency interferomeric radio telescope, operating in the benign radio frequency environment of remote Western Australia.  The MWA is the low frequency precursor to the Square Kilometre Array (SKA) and is the first of three SKA precursors to be operational, supporting a varied science mission ranging from the attempted detection of the Epoch of Reionisation to the monitoring of solar flares and space weather.  In this paper we explore the possibility that the MWA can be used for the purposes of Space Situational Awareness (SSA).  In particular we propose that the MWA can be used as an element of a passive radar facility operating in the frequency range 87.5 - 108 MHz (the commercial FM broadcast band).  In this scenario the MWA can be considered the receiving element in a bi-static radar configuration, with FM broadcast stations serving as non-cooperative transmitters.  The FM broadcasts propagate into space, are reflected off debris in Earth orbit, and are received at the MWA.  The imaging capabilities of the MWA can be used to simultaneously detect multiple pieces of space debris, image their positions on the sky as a function of time, and provide tracking data that can be used to determine orbital parameters.  Such a capability would be a valuable addition to Australian and global SSA assets, in terms of southern and eastern hemispheric coverage.  We provide a feasibility assessment of this proposal, based on simple calculations and electromagnetic simulations that shows the detection of sub-metre size debris should be possible (debris radius of $>0.5$ m to $\sim$1000 km altitude).  We also present a proof-of-concept set of observations that demonstrate the feasibility of the proposal, based on the detection and tracking of the International Space Station via reflected FM broadcast signals originating in south-west Western Australia.  These observations broadly validate our calculations and simulations.  We discuss some significant challenges that need to be addressed in order to turn the feasible concept into a robust operational capability for SSA.  The aggregate received power due to reflections off space debris in the FM band is equivalent to a $<1$ mJy increase in the background confusion noise for the long integrations needed for Epoch of Reionisation experiments, which is insignificant.
\end{abstract}


\section{Introduction}
Space debris consists of a range of human-made objects in a variety of orbits around the Earth, representing the remnants of payloads and payload delivery systems accumulated over a number of decades.  With the increasing accumulation of debris in Earth orbit, the chance of collisions between the debris increases (causing an increase in the number of debris fragments).  More importantly, the chance of collisions between debris and active satellites increases, posing a risk of damage to these expensive and strategically important assets.  This risk motivates the need to obtain better information on the population characteristics of debris (distribution of sizes and masses, distribution of orbits etc).  

Research into space debris is considered a critical activity world-wide and is the subject of the Inter-Agency Space Debris Coordination Committee\footnote{http://www.iadc-online.org/}, previously described in detail in many volumes, including a United Nations Technical Report on Space Debris \citep{UNreport} and a report from the Committee on the Peaceful Uses of Outer Space \citep{demp12}.  The American Astronomical Society maintains a Committee on Light Pollution, Radio Interference and Space Debris\footnote{http://aas.org/comms/committee-light-pollution-radio-interference-and-space-debris}, since space debris poses a risk to important space-based astrophysical observatories.

The risks posed by space debris, and the difficulties inherent in tracking observations and orbit predictions, were illustrated starkly on the 10$^{\rm th}$ of February 2009, when the defunct Russian Kosmos-2251 satellite collided with the active Iridium-33 satellite at a relative speed of over 11 km/s, destroying the Iridium satellite\footnote{Orbital Debris Quarterly News, 2009, NASA Orbital Debris Program Office, 13 (2)}.  Even with the global efforts to track space debris and predict their orbits, this collision between two satellites of 900 kg and 560 kg, respectively, and of $\sim$10 m$^{2}$ debris size, was unanticipated.  Further exploration of SSA capabilities is required to continue to minimise collision risks and provide better early warnings following collision breakups that multiply debris numbers and randomise orbits.

Methods for obtaining data on space debris include ground-based (radar observations and optical observations) and space-based measurements.  Previously, large ground-based radio telescopes (whose primary operations are for radio astronomy) have been used in a limited fashion to track space debris.  For example, the 100 m Effelsberg telescope in Germany, outfitted with a seven-beam 1.4 GHz receiver, was used for space debris tracking using reflected radiation generated by a high power transmitter \citep{ruitz05}.  Recently, a new method of ground-based space debris observation has been trialed, using interferometric radio telescope arrays (also primarily operated for radio astronomy), in particular using the Allen Telescope Array \citep{wel09}.  This technique utilises stray radio frequency emissions originating from the Earth that reflect off space debris and are received and imaged at the interferometric array with high angular resolution.  Scenarios such as this, with a passive receiver and a non-cooperative transmitter, are described as “passive radar”, a sub-class of the bi-static radar technique (transmitter and receiver at different locations).

This paper explores the possibilities offered by this technique, using a new low frequency radio telescope that has been built in Western Australia, the Murchison Widefield Array (MWA).  The MWA is fully described in \citet{tin13}.  Briefly, the MWA operates over a frequency range of 80 - 300 MHz with an instantaneous bandwidth of 30.72 MHz, has a very wide field of view ($\sim$2400 deg$^{2}$ at the lower end of the band), and reasonable angular resolution ($\sim$6 arcmin at the lower end of the band).

Recent observations with the MWA by \citet{mck12} have shown that terrestrial FM transmissions (between 87.5 and 108.0 MHz) reflected by the Moon produce a significant signal strength at the MWA.  \citet{mck12} estimate that the Equivalent Isotropic Power (EIP) of the Earth in the FM band is approximately 77 MW.  The ensemble of space debris is illuminated by this aggregate FM signal and will reflect some portion of the signal back to Earth, where the MWA can receive the reflected signals and form images of the space debris, tracking their positions on the sky as they traverse their orbits.

It should be noted that this technique makes use of the global distribution of FM radio transmitters, but that the technique is likely to work most effectively with receiving telescopes that are well separated from the transmitters.  If the receiving array is in close proximity to an FM transmitter, the emissions directly received from the transmitter will be many orders of magnitude stronger than the signals reflected from the space debris.  Therefore, the technique can likely only be contemplated using telescopes like the MWA, purposefully sited at locations such as the radio-quiet Murchison Radio-astronomy Observatory (MRO), located in Murchison Shire, 700 km north of Perth in Western Australia \citep{jon08}.  The MWA is a science and engineering Precursor for the much larger and more sensitive low frequency component of the Square Kilometre Array (SKA) \citep{dew09}, which will also be located at the MRO and is currently under development by the international SKA Organisation.

In this paper we estimate the conditions under which the MWA can detect space debris, in passive radar mode in the FM band.  In section 2 we consider a simple calculation of the expected signal strength at the MWA for one idealised scenario.  We also present comprehensive but idealised electromagnetic simulations that agree well with the simple calculation and show that space debris detection is feasible with the MWA.  For normal MWA observation modes, the detection of space debris of order $\sim$0.5 m radius and larger appears feasible.  If efforts are made to modify standard MWA observation modes and data processing techniques, then substantially smaller debris could be detected, down to $\sim$0.2 m in radius in low Earth orbits.  In section 3 we present observational tests that show the basic technique to be feasible, but also illustrate challenges that must be overcome for future observations.  In section 4 we discuss the utility of the technique, complementarity with other techniques, and suggest future directions for this work.

\section{Detectability of space debris with the MWA}
As an example that illustrates the feasibility of this technique, we
consider the MWA as the receiving element in a passive radar system in
the FM band and consider a single transmitter based in Perth, Western
Australia (specifically the transmitter for call sign 6JJJ: Australian Communications and Media Authority [ACMA]
Licence Number 1198502).  The transmitter is located at
(LAT,LONG)=($-32\degr00\arcmin42\arcsec$,
  $116\degr04\arcmin58\arcsec$) and transmits a mixed polarisation
signal, with an omnidirectional radiation pattern in the horizontal plane, at 99.3 MHz over the FM
emission standard 50 kHz bandwidth and with an Effective Radiated
Power (ERP) of 100 kW.  The MWA, acting as receiver, is located at
(LAT,LONG)=($-26\degr42\arcmin12\arcsec$,
  $116\degr40\arcmin15\arcsec$), $\sim$670 km from the
transmitter.  

We consider an idealised piece of space debris (a
perfectly conducting sphere of radius $r$, in metres) at a distance 
of $R_{r}$ km 
from the MWA site.  We
denote the distance between the transmitter and the space debris as
$R_{t}$ km.  For simplicity, we model the
Radar Cross Section (RCS, denoted by $\sigma$) of the debris as the
ideal backscatter from a conducting sphere in one of two domains that
are approximations of Mie scattering \citep{str41}: 1) when the wavelength
$\lambda>2\pi r$, the RCS is described by Rayleigh scattering and
$\sigma= 9\pi r^{2}(\frac{2 \pi r}{\lambda})^{4}$; and 2) when
$\lambda<<2\pi r$, $\sigma=\pi r^{2}$, corresponding to
  the geometrical scattering limit.  The greatest interest is in
case 1, as the space debris size distribution is dominated by small
objects (although the example given in the introduction shows that
even the largest items of space debris pose serious unknown risks).
For case 1 we obtain Equation 1, below, following the well-known
formula for bi-static radar \citep{wills05},

\begin{equation}
S=3.5\times10^{6}\frac{P_{t} G r^{6} \nu^{4}}{R_{t}^{2} R_{r}^{2} B}
\label{eqn:flux}
\end{equation}

where S is the spectral flux density of the received signal at the MWA in astronomical units of Jansky (1 Jy = 10$^{-26}$ Wm$^{-2}$Hz$^{-1}$), $P_{t}$ is the ERP in kW, $G$ is the transmitter gain in the direction of the space debris, $B$ is the bandwidth in kHz over which the signal is transmitted, $R_{t}$, $R_{r}$ and $r$ are as described above, and $\nu$ is the center frequency of the transmitted signal in MHz.

For $P_{t}$=100 kW, $R_{r}$=1000 km, $R_{t}=$1200 km, $G$=0.5, $B$=50 kHz, $r$=0.5 and $\nu$=99.3 MHz, the spectral flux density, $S$, is approximately 4 Jy.  The assumption of $G$=0.5 is based on an idealised dipole antenna transmitter radiation pattern which is omni-directional in the horizontal plane.  FM transmitter antenna geometries vary significantly, with local requirements on transmission coverage dictating the directionality and thus the antenna geometry.  Phased arrays can be used to compress the radiation pattern in elevation and different antenna types can produce gain in preferred azimuthal directions.  Therefore, in reality, significant departures from a simple dipole is normal.  Propagation losses in signal strength due to two passages through the atmosphere are assumed to be negligible.  Due to the approximations described above, this estimate should be considered as order-of-magnitude only.  However, the calculation indicates that the detection of objects of order 1 m in size is feasible.  In a 1 second integration with the MWA over a 50 kHz bandwidth, an image can be produced with a pixel RMS of approximately 1 Jy \citep{tin13}, making the reflected signal from the space debris a several sigma detection in one second.

According to equation 1, the MWA will be sensitive to sub-metre scale space debris.  However, for FM wavelengths, sub-metre scale debris are in the Rayleigh scattering regime, meaning that the RCS drops sharply for smaller debris, dramatically reducing the reflected and received power.  In order to more systematically explore the detectibility of debris of different sizes and relative distances between transmitter, debris, and receiver, we have performed a series of electromagnetic simulations.

The simulations performed take the basic form described above, in
which a single FM transmitter (idealised dipole radiation pattern for the transmitter, omni-directional in the horizontal plane) at a given location was considered
(separate simulations were performed for 11 different transmitters
located in the south-west of Western Australia), along with the known
location of the MWA as receiving station.  The debris were modelled as being directly above the MWA.  A
range of different debris altitudes was modeled: 200 km; 400 km; 800
km; and 1600 km (as for the calculation above, propagation losses due to the atmosphere are considered negligible).  The debris was assumed to be a perfectly conducting
sphere, with a range of radii: 0.2 m; 0.5 m; 1 m; and 10 m.  The
simulations proceeded using the {\it XFdtd} code from Remcom
Inc\footnote{http://www.remcom.com/xf7}.  {\it XFdtd} is a
general-purpose electromagnetics analysis code based on the
finite-difference, time-domain technique and can model objects of
arbitrary size, shape and composition.  The $11$ (transmitters) $\times$ $4$
(radii) $\times$ $4$ (altitudes) $= 176$ runs of {\it XFdtd} were calculated
on a standard desktop computer over the course of several hours.  The
simulations compute output that includes the RCS according to the full
Mie scattering solution and the spectral flux density at the location
of the MWA, for each combination of debris size, altitude and
transmitter.

Figure 1 shows a selection of results from the simulations, for the Perth-based 6JJJ transmitter and the ranges of debris size and altitudes listed above.  Also shown is a single point representing the result of the simple calculation from Equation 1.  The simple calculation overestimates, by a factor of approximately two, relative to the {\it XFdtd} simulation, the flux density at the MWA.  This level of agreement is reasonable, given the simplifying assumptions made for Equation 1.  


\begin{figure}[ht]
\plotone{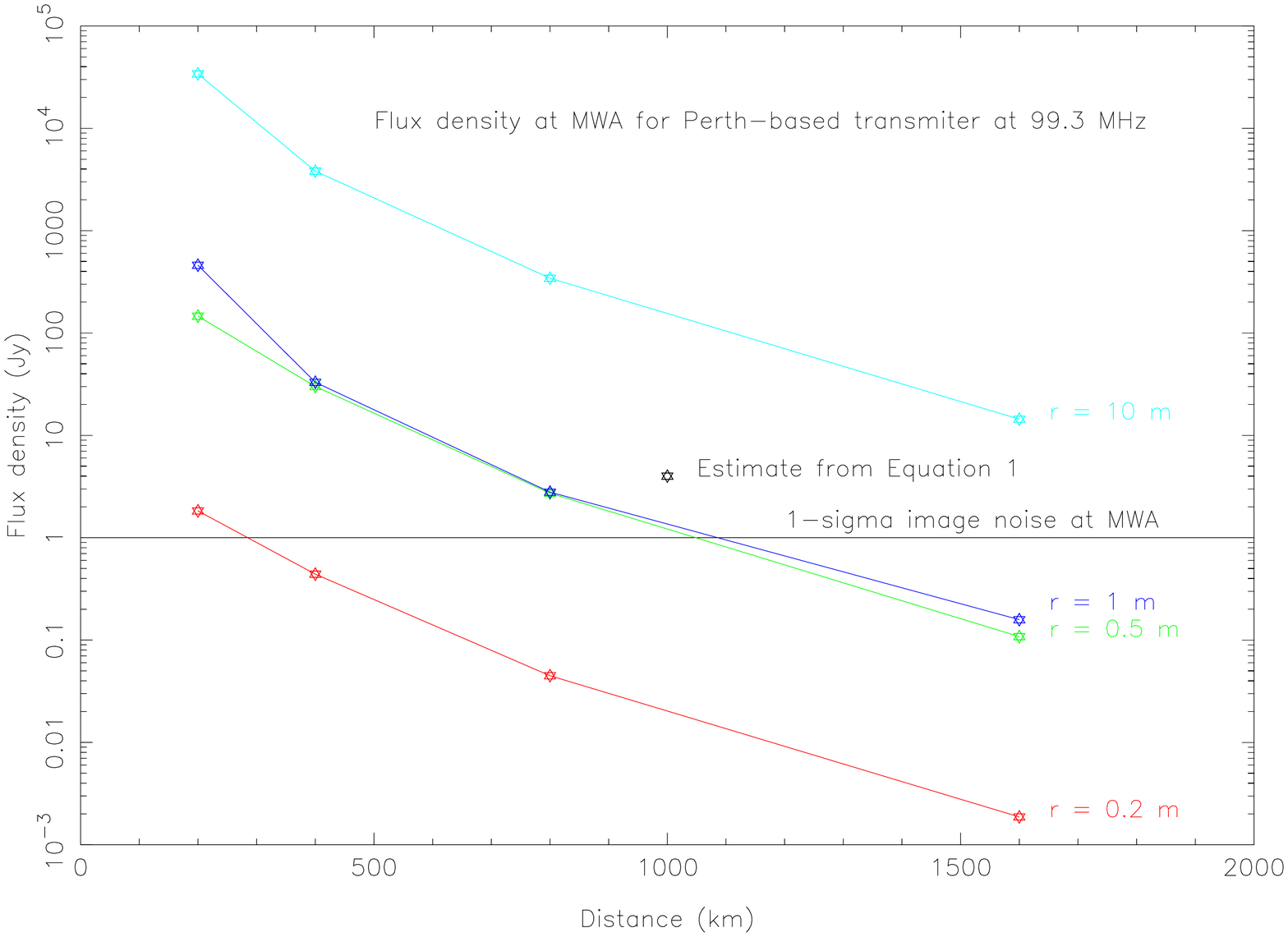}
\caption{The results of electromagnetic simulations, as described in the text.  These simulations model the flux density at the MWA for space debris modelled as a perfectly conducting sphere (0.2, 0.5, 1 and 10 m radii) at altitudes of 200, 400, 800 and 1600 km above the MWA site, with a transmitter at 99.3 MHz based near Perth ($\sim$700 km from the MWA).  Also shown is a single point representing the example calculation based on Equation 1, with $r=0.5$ m}
\end{figure}

Figure 1 shows that debris with $r >$ 0.5 m can plausibly be detected up to altitudes of approximately 800 km with a one second observation.  The solid line in Figure 1 denotes the one sigma sensitivity calculation made for the simple example, above, based on information from \citet{tin13}, showing that a three sigma detection can be made for debris of $r =$0.5 m at 800 km.  Of course, if longer observations can be utilised, the detection thresholds can be reduced.  For example, a ten second observation would yield a three sigma detection of debris with $r =$ 0.5 m at 1000 km.

The solution to the detection of smaller ($\sim$10 cm) scale debris with low frequency passive radar in the past has been to use ERPs of gigawatts, as with the 217 MHz NAVSPACECOM ``Space Fence'' \citep{pet12}.  An alternative approach is to build larger and more sensitive receiving antennas.  The natural evolution of the MWA is the much larger and more sensitive low frequency SKA, which will have a factor of $\sim$1000 more receiving area and be far more sensitive than the MWA.  The low frequency SKA may have great utility for space debris tracking later this decade.

\section{Verification of the technique}
\begin{figure}[ht]
\plottwo{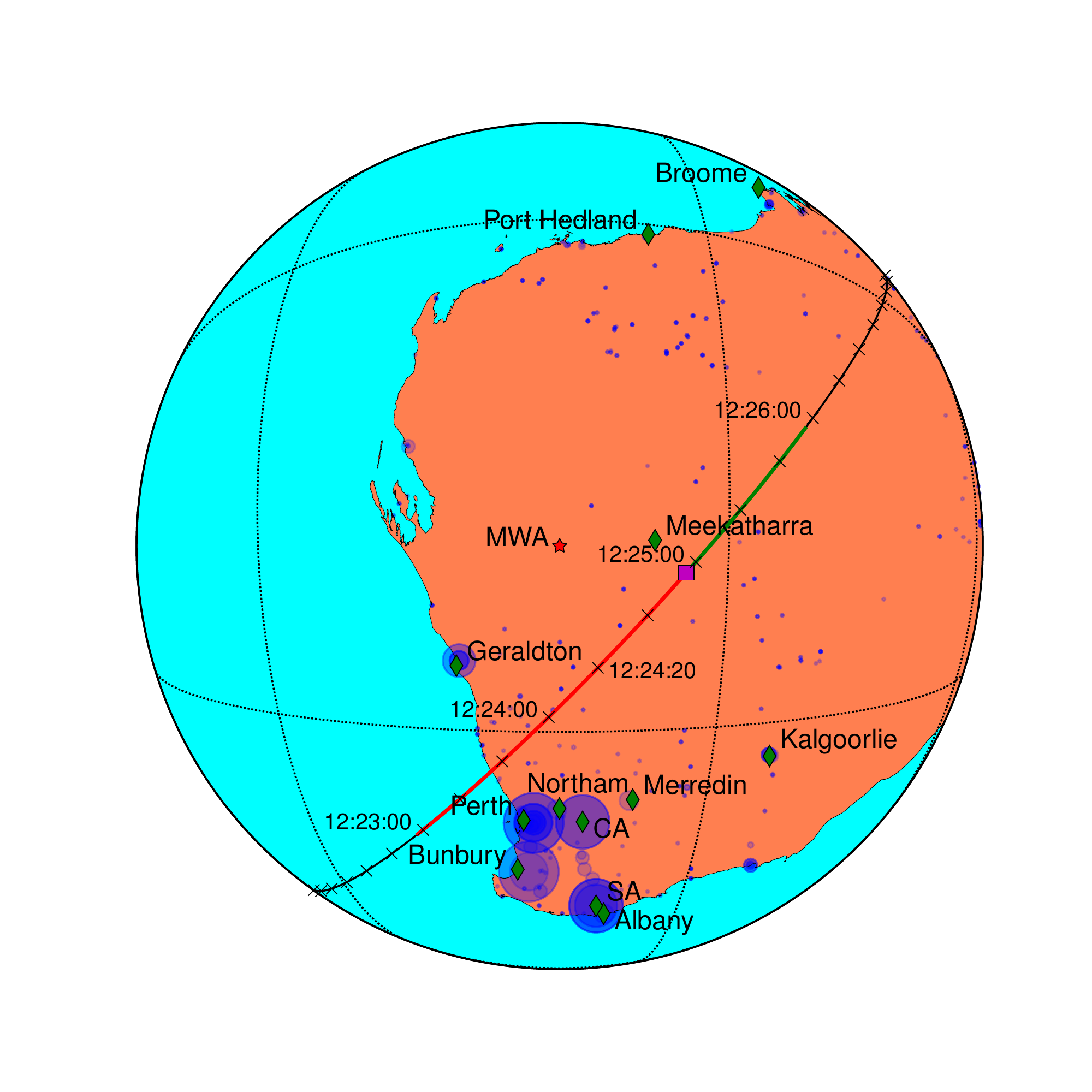}{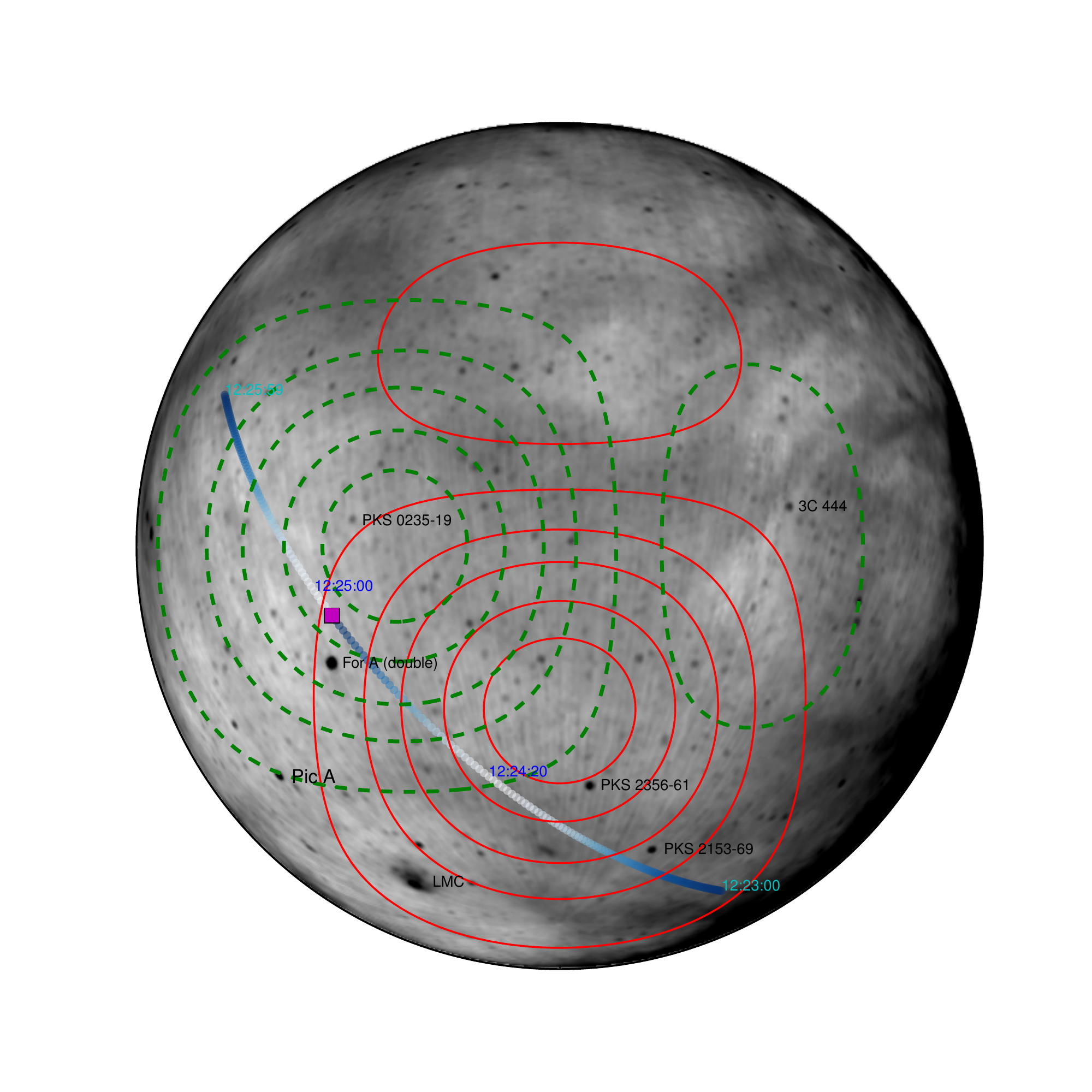}
\caption{Left: The track of the ISS over Western Australia on
  2012~November~26, between 12:23:00 and 12:26:00 UT.  The red and
  green segments of the track denote the time ranges for the two
  different pointed observations undertaken with the MWA, with ticks
  located every 20\,s and the square filled marker denoting the changeover time for two pointings.  Blue circles denote the location and power
  (proportional to diameter of circle, with largest circles
  representing 100 kW transmitter ERP) of FM transmitters.  Green
  diamonds denote the names of towns/regions associated with the
  transmitters.  The image is a projection representing the view of
  Australia from a low-Earth orbit satellite directly over the MWA
  site. Right: The track of the ISS plotted on the celestial sphere as
  visible from the MWA site.  The image is an orthographic projection
  of the \citet{has82} 408\,MHz map, with some individual radio
  sources labeled.  The red and green contours are contours of primary
  beam gain (at 1\%, 10\%, 25\%, 50\%, and 75\% of the peak gain) for
  the two pointed observations.  The blue points are the track of the
  ISS, with color from dark blue to white representing the primary
  beam gain from low to high and the square filled marker denoting the changeover time for two pointings.  North is up and east is to the left
  (note that this orientation is opposite that in the left-hand
  panel).}
\label{fig:track}
\end{figure}

We verified the technique outlined above with an observation using the
MWA during its commissioning phase.  This is a subarray of 32/128 tiles with a maximum east-west baseline of roughly
1\,km and a maximum north-south baseline of roughly 2\,km, giving a
synthesized beam of roughly $5\arcmin \times 10\arcmin$ at a
frequency of $\sim$100 MHz.  We observed an overflight of the
International Space Station (ISS) on 2012~November~26, as
shown in Figure~\ref{fig:track}.  The center frequency of our
30.72\,MHz bandpass was 103.4\,MHz, and we correlated the data with a
resolution of 1\,s in time and 10\,kHz in frequency.  We used two
different array pointings, as plotted in the right panel of
Figure~\ref{fig:track}.  The first was from 12:20:00\,UT to
12:24:56\,UT and was pointing at an azimuth of $180\degr$ (due south)
and an elevation of $60\degr$.  The second was from 12:24:56\,UT to
12:29:52\,UT at an azimuth of $90\degr$ (due east) and an elevation of
$60\degr$.  As discussed in \citet{tin13} and verified by
  \citet{wil12}, the MWA has an extremely broad primary beam (roughly
  $35\degr$ FWHM) which was
  held fixed during each observation.  We observed the overflight of
  the ISS between 12:23:00\,UT and 12:26:00\,UT, when it was at a
  primary beam gain of $>1$\%, traversing about $112\degr$ on the sky.

During the pass, the ISS ranged in distance between approximately
850\,km (at the low elevation limit of the MWA, approximately
30$^{\circ}$) and approximately 400\,km (near zenith at the MWA) (Figure 2); its angular speed across the sky is
approximately 0.5$^{\circ}$ per second, corresponding to approximately
five synthesised beams with this antenna configuration.

To image the data, we started with an observation pointed at the
bright radio galaxy 3C~444 taken before the ISS observations. This
observation was used as a calibration observation, determining complex
gains for each tile assuming a point-source model of
3C~444, which was sufficient to model the source given that its extent
($<4$\arcmin) is the same as the FWHM of the synthesised beam of the
MWA subarray at this frequency. While we changed pointings between
this observation and those of the ISS, we have previously found that the
instrumental phases are very stable across entire days, and that only
the overall amplitudes need to be adjusted for individual
observations.

Antenna-based gain solutions were obtained
using the \textsc{CASA} task \textit{bandpass} on a 2-minute
observation. Solutions were obtained separately for each 10\,kHz
channel, integrating over the full two minutes. The flux density used
for 3C~444 was 116.7\,Jy at the central frequency of 103\,MHz, with a
spectral index of $-0.88$. A S/N$>3$
was required for a successful per-channel, per-baseline solution, and
the overall computed gain solutions for each antenna were of S/N$>5$.
Calibration solutions were applied to the ISS observations using the \textsc{CASA} task
\textit{applycal} and the corrected data were written to
\textsc{uv-fits} format for imaging in \texttt{miriad} \citep{sau95}.

For the two ISS observations, we flagged bad data in individual
fine channels (10\,kHz wide) associated with the centers and edges of
our 1.28\,MHz coarse channels.  Because of the extremely fast motion
of the ISS, we created images in 1\,s intervals (the integration time
used by the correlator) using \texttt{miriad}.  We started by imaging the
whole primary beam ($40\degr \times 40\degr$), separating the
data into $3\times 10.24$\,MHz bandpasses.  The bottom two bandpasses covered the FM band,
while the top bandpass was above the FM band. We used a
$1\arcmin$ cell size and cleaned emission from Fornax~A.  Note that
\texttt{miriad} does not properly implement wide-field imaging, but our
images used a slant-orthographic projection such that the synthesized
beam is constant over the image \citep{ord10}.

\begin{figure}[ht]
\plotone{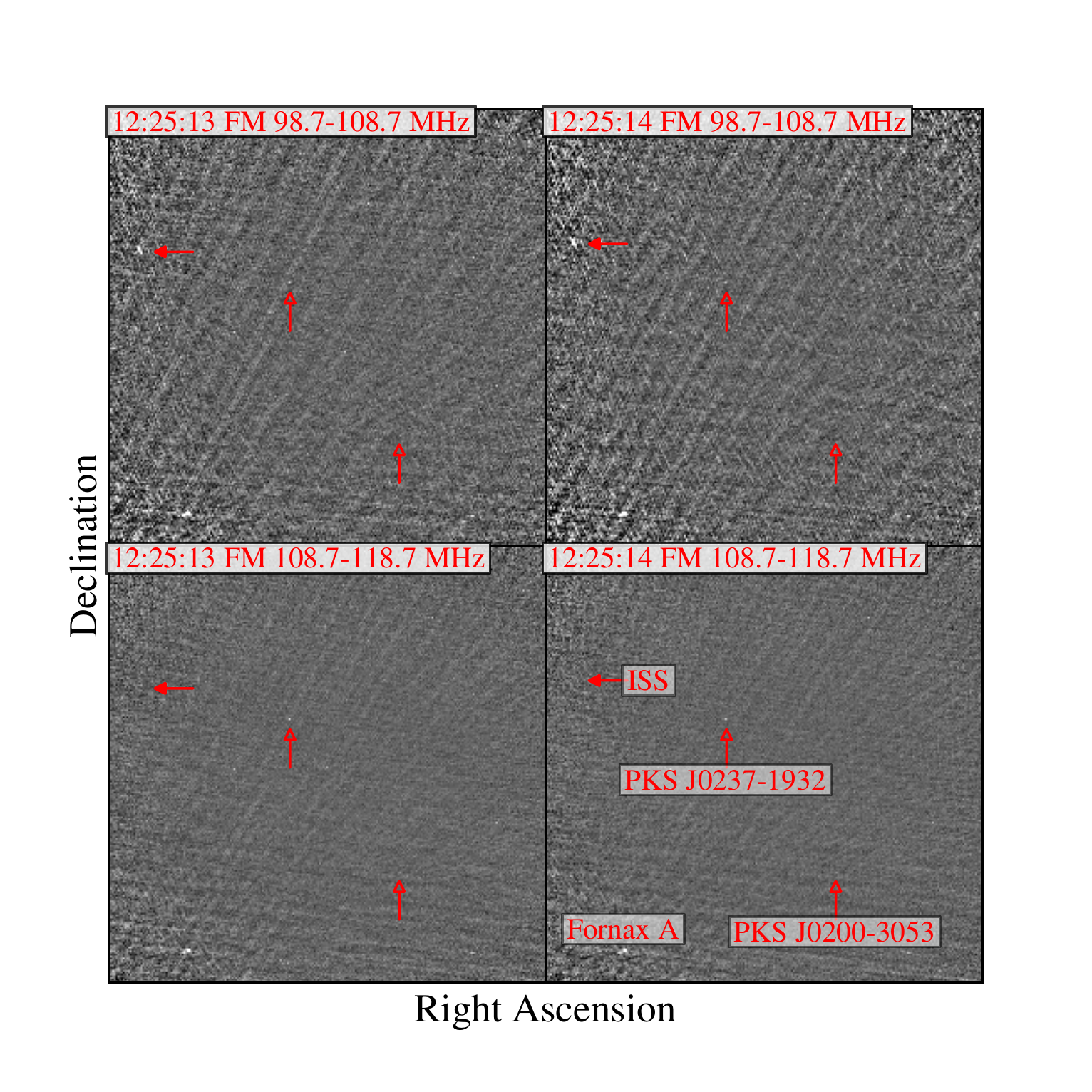}
\caption{Images generated for the ISS overflight.  The left panels
  show the 1\,s interval starting at 12:25:13\,UT, while the right
  panels show the interval starting 1\,s later at 12:25:14\,UT.  The
  top panels show the 10\,MHz covering the top end of the FM band
  (98.7\,MHz--108.7\,MHz) while the bottom panels show the 10\,MHz
  above that, which does not include any of the FM band.  The
  double-lobed radio source Fornax\,A as well as the radio galaxies
  PKS~J0200$-$3053 and PKS~J0237$-$1932 are labeled (with arrows).  The predicted position of the ISS at the two times is indicated; it
  is readily visible as a streak in the top images, but is not visible in the bottom images.
  The greyscale scales linearly with flux, and are the same for all the panels.  The images have been
  corrected for primary beam gain, which is why the noise appears to
  increase toward the left hand side.  North is up, east to the left, and
  these images cover roughly $35\degr$ on a side.  10 MHz bands have been used, even though the FM signals only occupy a fraction of the 98.7 -- 108.7 MHz band, so that the radio galaxies can be detected.
}
\label{fig:images}
\end{figure}

The results covering the two time intervals and two bandpasses are shown in
Figure~\ref{fig:images}. The ISS is readily visible as a streak moving
through our images, but is only visible in the images that cover the
FM band as would be expected from reflected terrestrial FM emission
(as in \citealt{mck12}).  We predicted the position of the ISS based
on its two-line ephemeris (TLE) using the \texttt{pyephem} package\footnote{http://rhodesmill.org/pyephem}.  The predicted position agrees with the observed position to within the observational uncertainties.

\begin{figure}[ht]
\plotone{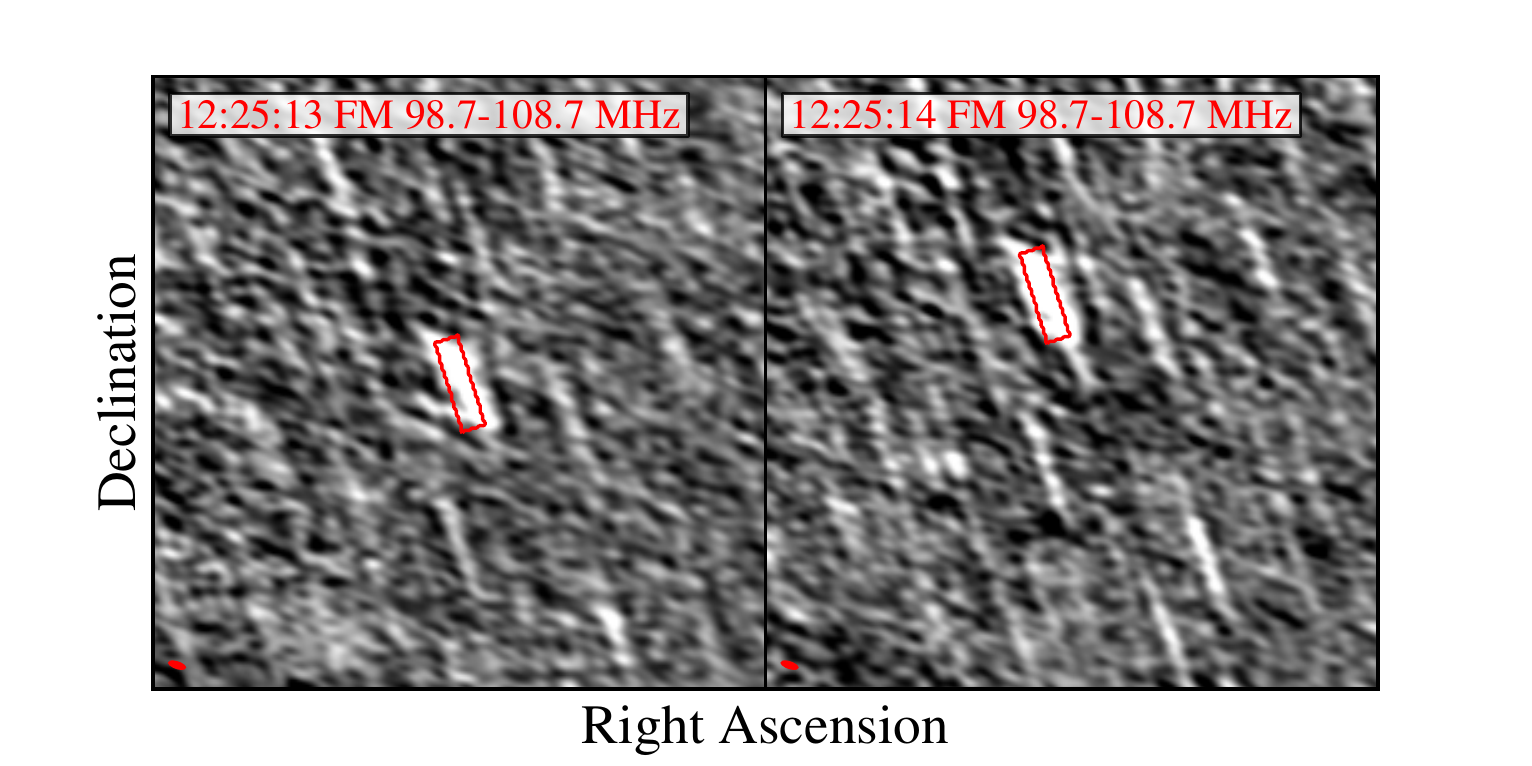}
\caption{A zoom in of the image of the ISS in the FM band over a 1
  second integration period.  The left panel shows the 1\,s interval
  starting at 12:25:13\,UT, while the right panel shows the interval
  starting 1\,s later at 12:25:14\,UT (same times and data as shown in \ref{fig:images}). These images cover the
  98.7--108.8\,MHz bandpass and are centered on the position of the
  ISS at 12:25:13\,UT.  The red boxes show the region over which the
  flux was summed to create the dynamic spectra in
  Figure~\ref{fig:dynspec}.  North is up, east to the left, and these
  images cover roughly $4\degr$ on a side.  The synthesised beam is shown in the bottom left hand corner of the images.}
\label{fig:zoom}
\end{figure}

\begin{figure}[ht]
\epsscale{0.7}
\plotone{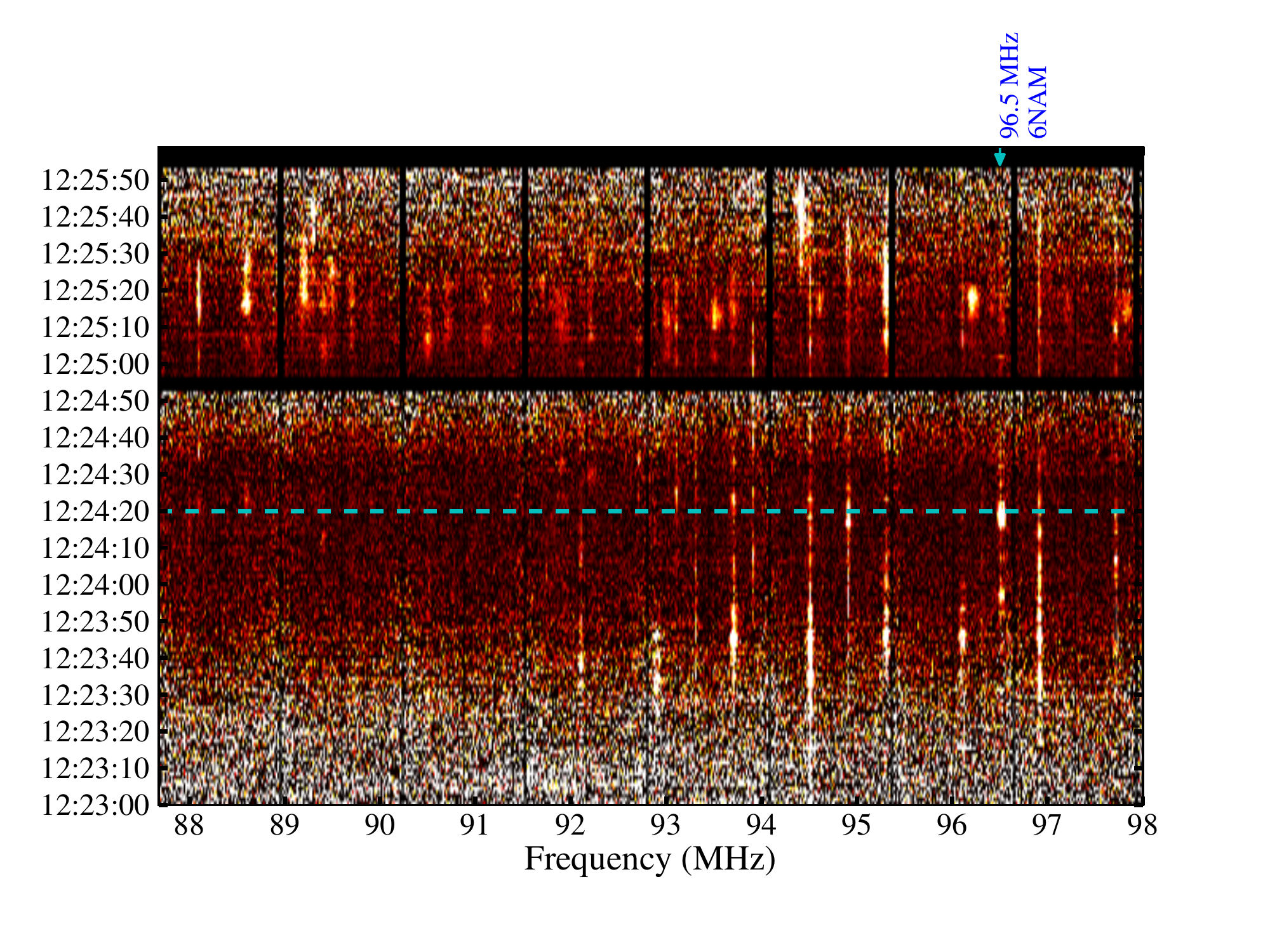}\\
\plotone{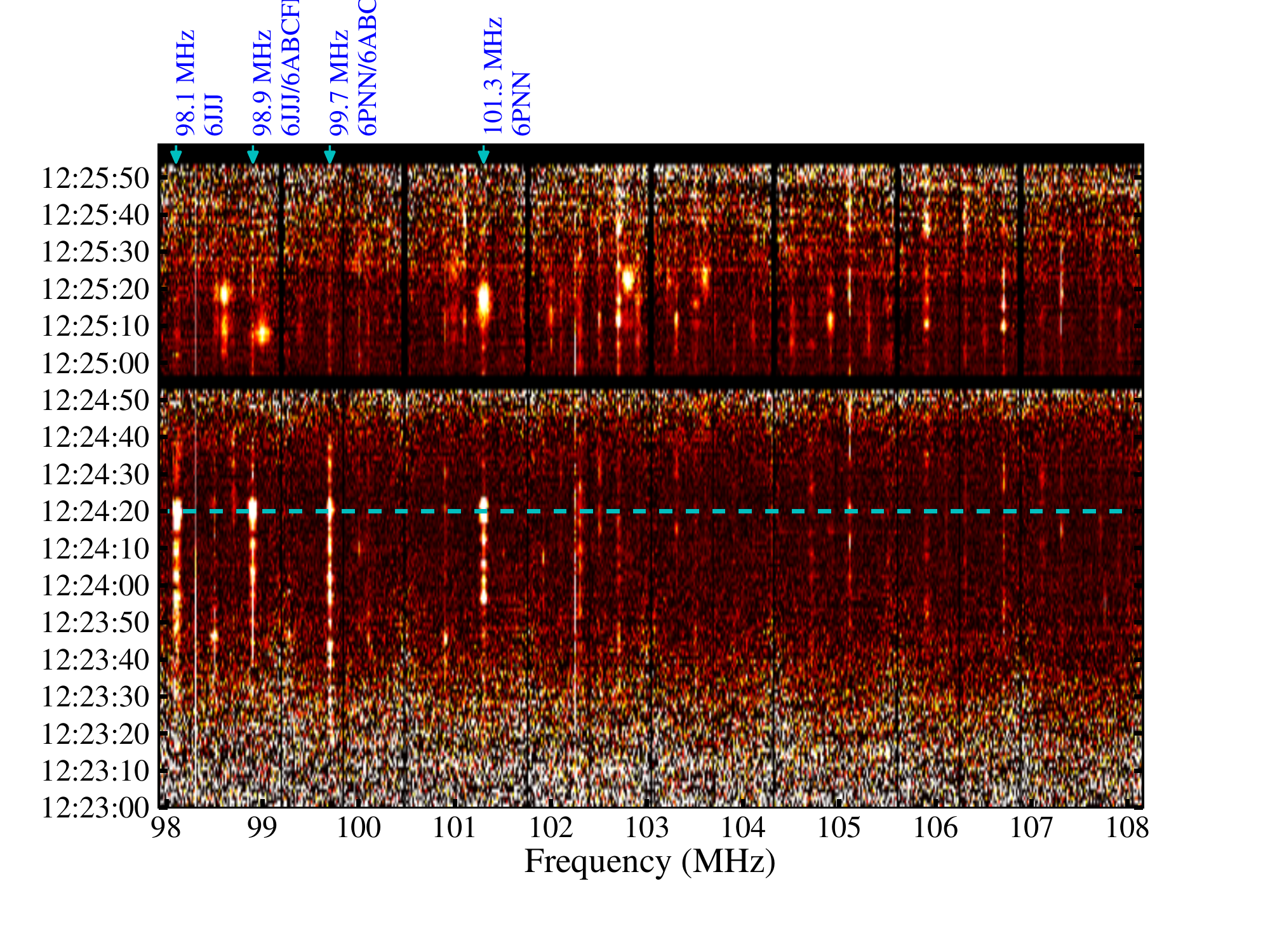}
\caption{Dynamic spectra over the time range 12:23:00 to 12:26:00 UT
  and over the frequency ranges 88--98\,MHz (top panel) and
  98--108\,MHz (bottom panel).  The black regions are those with no
  data, either because of the gap between the observations or the
  individual channels flagged during imaging.  The color scale in both
  images is the same and is linearly proportional to flux density, with the white stretch indicating the strongest signals.
  The specific time discussed in the text, 12:24:20\,UT, is marked
  with a dashed line. Individual FM frequencies from
  Table~\ref{tab:transmit} are also labeled.  The noisier portions at
  the beginning and end of the first observation and at the end of the
  second observation are because the flux densities have been
  corrected for the primary beam gain, which is low at those times.}
\label{fig:dynspec}
\end{figure}

To determine more quantitatively the nature of the emission seen in
Figure~\ref{fig:images}, we imaged each 1\,s interval and each 10\,kHz
fine channel separately (as before, this was done in \texttt{miriad}).
We did not deconvolve the images at all; most of them had no signal or
the emission from the ISS was very weak.  The emission from the ISS is
not a point source (smeared due to its motion) and adding deconvolution for a non-stationary object to our already
computationally-demanding task was not feasible.  In each of our $180
\times 3072$ dirty images, we located the position of the ISS based on
its ephemeris.  We then measured its flux density by adding up the
image data over a rectangular region of $10\arcmin$ (comparable to our
instrumental resolution) in width and with a length appropriate for
the instantaneous speed of the ISS (up to $40\arcmin$ in a 1\,s
interval), as seen in Figure~\ref{fig:zoom}.  The resulting dynamic spectra
showing flux density as functions of time and frequency are shown in
Figure~\ref{fig:dynspec}.  These spectra have been corrected for the
varying primary beam gain of the MWA over its field-of-view based on
measurements of the sources PKS~2356$-$61 (first observation, assuming
a flux density of $166\,$Jy at 97.7\,MHz) and PKS~J0237$-$1932 (second
observation, assuming a flux density of 35\,Jy at 97.7\,MHz); we have
ignored the frequency-dependence of the primary beam gain for the
purposes of this calculation.  Note that the emission seen in
Figure~\ref{fig:dynspec} is manifestly \textit{not} radio-frequency
interference (RFI) that is observed as a common-mode signal by all of
our tiles.  It is only visible at a discrete point in our synthesized
images corresponding to the position of the ISS
(Figure~\ref{fig:zoom}).  We ran the same routines used to create the
dynamic spectra in Figure~\ref{fig:dynspec} but using a position offset from the ISS
position, and we see almost no emission, with what we do
see consistent with sidelobes produced by bright emission in
individual channels, due to the ISS.

The ISS has an approximate maximum projected area of $\sim$1400 m$^{2}$, is of mixed composition (thus not well approximated as perfectly conducting), has a highly complex geometry, and was at an unknown orientation relative to the transmitter(s) and receiver during the observations.  The detailed transmitter antenna geometries and gain patterns are not known.  Thus, it is very difficult to accurately predict the flux density we would expect from either simple calculations or electromagnetic simulations.  

We can, however, unambiguously identify the origin of some of the transmissions reflected off the ISS.  For example, we consider the dynamic spectrum from our observations in Figure 5 at the time 12:24:20 UT.  At this time there are clearly a number of strong signals detected, plus a large number of much weaker signals.  The five strongest signals at this time, in terms of the integrated fluxes (Jy.MHz, measured by summing over individual transmitter bands in the dynamic spectra to a threshold defined by the point at which the derivative of the amplitude changed sign), are listed in Table 1, along with transmitters that can be identified as the origin of the FM broadcasts \citep{acma}.  Each of the five identified transmitters are local to Western Australia and, as can be seen from Figure 2, are all relatively close to the ISS at this time.  Each of the five transmitters have omni-directional antenna radiation patterns in the horizontal plane.

If we take the reported ERP for a typical station from
  Table~\ref{tab:transmit} and use
  it with Equation~\ref{eqn:flux} to estimate the received flux, we
  find values of roughly $10^4\,{\rm Jy.MHz}$, assuming the full
  $1400\,{\rm m}^2$ of reflecting area for the ISS.  Given its varying
  composition and orientation, the measured fluxes in
  Table~\ref{tab:transmit} are reasonably consistent with this estimate.

\begin{table}[ht]
  \begin{tabular}{ c | c c r r c c c} \hline
    \#   & Freq.  & Flux      & Call sign(s) & Location(s) & ERP(s) &$R_{t}$ & $R_{r}$  \\
         & (MHz)  & (Jy.MHz)  &           &          & (kW) & (km) & (km)\\ \hline
    1    &98.1&3055&6JJJ&Central Agricultural&80&537 & 489 \\
    2    &96.5&1907&6NAM&Northam&10&518 & 489 \\
    3    &98.9&1799&6ABCFM/&Central Agricultural/&80/10&537/498&489 \\
         &    &    &6JJJ   &Geraldton&&& \\
    4    &99.7&1718&6ABCFM/&Central Agricultural/&80/10&537/498&489 \\
         &    &    &   6PNN&Geraldton&&& \\
    5    &101.3&1672&6PNN&Geraldton&10&498&489 \\ \hline
 \end{tabular}
   \caption{The five strongest detected signals at 12:24:20 UT and transmitters identified with those signals\label{tab:transmit}}
\end{table}

Many, but not all, of the much weaker signals detected can also be
identified.  However, it is clear that some of these weaker signals
are from powerful transmitters at much larger distances, making unique
identification of the transmitter more difficult (FM broadcasts are
made on the same frequencies in different regions).  Furthermore, it
is clear that transmitters local to Western Australia, and of
comparable power to those listed in Table 1, have not been detected in
our observations.  This is likely to be the result of the complex
transmitter/reflector/receiver geometries mentioned above and the
rapid evolution of the geometry with time, illustrated by the rapid
evolution of the received signals identified in Table 1.  The five
strongest signals at 12:24:20 UT are strongest over a 5 second period,
have a highly modulated amplitude over a further 30 second period, but
are generally very weak or not detected over the majority of our
observation.  Thus, it is highly likely that other transmitters of
comparable strength would be detected at other times, when favourable
geometries prevail.  At 12:24:20 UT, it appears that the geometries
were favourable for the Central Agricultural, Northam and Geraldton
transmitters, but not for transmitters of comparable power in Perth,
Bunbury and Southern Agricultural.  We note that, while the strongest signals detected originated from transmitters with omni-directional transmitting antennas (see above), the transmittors of comparable power not detected (Perth, Southern Agricultural and Bunbury) also have omni-directional antennas.  Thus, the most likely factor driving detection or non-detection is the transmitter/reflector/receiver geometry, rather than transmitting antenna directionality.

The modulation in power of the reflected radiation is likely at least partly due to ``glints'' from the large, complex structure, which creates a reflected radiation pattern whose finest angular angular scale is $\sim \lambda/d$, where $\lambda$ is the wavelength of the radiation and $d$ is the extent of the object. For $d\sim100$ m (full extent of the ISS) and $\lambda \sim 3$ m, the glints are  $>2^{\circ}$ in angular extent. At an altitude of $\sim$500 km, the smallest glint footprint on the surface of the Earth is $\sim10$ km. With an ISS orbital speed of 8 kms$^{-1}$, the glint duration at the MWA is of order one second. This corresponds well to some of the modulation timescales seen in Figure 5, although modulation exists on longer timescales (10 seconds and longer). These longer timescales may be due to glints involving structures smaller than the full extent of the ISS (single or multiple solar panels, for example).  A full analysis of the modulation structure of the signals is extremely complex and cannot be performed at a sufficiently sophisticated level to be useful at present.  In the future, detailed electromagnetic simulations of complex reflector geometries may be used to gain insight into the modulations, but such an analysis is beyond the scope of the current work.

In long integrations, such as required for Epoch of Reionisation experiments \citep{bow13}, the effect of having 10 pieces of space debris in the MWA primary beam at any given time with flux densities $>$1 Jy is the equivalent of a $<$1 mJy additional confusion noise foreground component in the FM band.  This contribution will have no discernable impact on the MWA's science goals for observations in the FM band.

\section{Discussion and future directions}
At metre size scales, of order 1000 pieces of debris are currently known and tracked.  On average, up to approximately 10 such pieces will be present within the MWA field-of-view at any given time, allowing continuous, simultaneous detection searches and tracking opportunities.  

These estimated detection rates naturally lead to two primary capabilities.  The first is a blind survey for currently unknown space debris, provided by the very large instantaneous field-of-view of the MWA.  The second capability is the high cadence detection and tracking of known space debris.  The wide field-of-view of the MWA means that large numbers of debris can be simultaneously detected and tracked, for a substantial fraction of the time that they appear above the local horizon.

Imaging with the MWA can provide measurements of the right ascension and declination (or azimuth and elevation) of the space debris, as seen from the position of the MWA (convertable to topocentric right ascension and declination).  The MWA will also be able to measure the time derivatives of right ascension and declination, giving a four vector that defines the ``optical attributable''.  Measurements of the four vector at two different epochs allow an estimate of the six parameters required to describe an orbit.  \citet{farn09} describe methods for orbit reconstruction using the optical attributable, taking into account the correlation problem (being able to determine that two measurements of the four vector at substantially different times can be attributed to the same piece of space debris).

The calculations and practical demonstrations presented above show that the basic technique using the MWA is feasible and worthy of further consideration as a possible addition to Australia's SSA capabilities, particularly considering that the MWA is the Precursor to the much larger and more sensitive low frequency SKA.  However, the calculations and observational tests presented here also point to the challenges that will need to be met in order to develop this concept into an operational capability.

Overall, these challenges relate to the non-standard nature of the imaging problem for space debris detection, compared to the traditional approach for astronomy.  In radio interferomeric imaging, a fundamental assumption is made that the structure of the object on the sky does not change over the course of the observation.  This assumption is broken in spectacular fashion when imaging space debris, due to their rapid angular motion across the sky relative to background celestial sources and due to the variation in RCS caused by the rapidly changing transmitter/reflector/receiver geometry.

Another standard assumption used in astronomical interferometric imaging is that the objects being imaged lie in the far field of the array of receiving antennas; the wavefronts arriving at the array can be closely approximated as planar.  It transpires that for space debris, most objects lie in the transition zone between near field and far field at low radio frequencies, for an array the size of the MWA.  Signal to noise is degraded somewhat when the standard far-field assumption is adopted, as an additional smearing of the imaged objects results.  It is worth noting that this will be a much larger effect for the SKA, with a substantially larger array footprint on the ground.  Thus, accounting for this effect for the MWA will be an important step toward using the SKA for SSA purposes.  For example, for an object at 1000 km and the MWA spatial extent of 3 km, the deviation from the plane wave assumption translates into more than a radian of phase error across the array at a frequency of 100 MHz, which produces appreciable smearing of the reconstructed signal with traditional imaging.  It may be possible to use the near field nature of the problem to undertake 3-dimensional imaging of the debris, using the wavefront curvature at the array to estimate the distance to the debris.

Further significant work is required to implement modified signal
and image processing schemes for MWA data that take account of these
time-dependent and geometrical effects and sharpen the detections
under these conditions.  For example, the flexible approach to
producing MWA visibility data (the data from which MWA images are
produced) allows modifications to the signal processing chain to
incorporate positional tracking of an object in motion in real-time.
Additionally, it is possible to collect data from the MWA in an even
more basic form (voltages captured from each antenna) and apply high
performance computing in an offline mode to account for objects in
motion.  This would also enable measurement of the Doppler
  shifts of individual signals (expected to be $\sim {\rm few}\,$kHz,
  which is smaller than our current 10\,kHz resolution) that could
  enable separation of different transmitters at the same frequency
  and unambiguous identification of transmitters (a Doppler shift
  pattern fixes a one-dimensional locus perpendicular to the path of
  the reflecting object, so combining two passes allows
  two-dimensional localizations).  That in turn would help with
  constraining the basic properties of the reflecting objects\footnote{Given the changing geometry between passes, more than two might be required before a given transmitter is identified more than once with sufficient signal-to-noise for localization.  Conversely, if the Doppler measurement is sufficiently significant and the transmitter is sufficiently isolated, a single one-dimensional localization might be enough for unique identification.  We intend to test these ideas with future observations.}.
Finally, modifications to image processing algorithms can be applied
in post-processing to correct for near-field effects, essentially a
limited approximation of the same algorithms that can be applied
earlier in the signal chain.  The results of this future work will be
reported in subsequent papers.

Once these improvements are addressed, the remaining significant challenges revolve around searching for and detecting signals from known and unknown space debris.  These challenges are highly aligned with similar technical requirements for astronomical applications, in searches for transient and variable objects of an astrophysical nature.  A large amount of effort has already been expended in this area by the MWA project, as one of the four major science themes the project will address \citep{bow13,mur13}.  To illustrate the challenges, our flux estimates require (for example) greater
  than 4$\sigma$ significance, but that is for a single trial.  This
  is correct when looking to recover debris with an approximately
  known ephemeris, but blind searches have large trial factors that
  can require significantly revised thresholds. Given
  the full $2400\,{\rm deg}^2$ field-of-view of the MWA, we must
  search over $\sim 10^4$ individual positions in each image to
  identify unknown debris.  Combined with 3000 channels and 3600 time
  samples (for an hour of observing; note that we might observe with a
  shorter integration time in the future) this results in $10^{11}$
  total trials.  So our 4$\sigma$ threshold (which implies 1 false
  signal in $10^4$ trials) would generate $10^7$ false signals, and we
  would need something like an $8\sigma$ threshold to be assured of a
  true signal.  Given the steep dependence of flux on object size
  (Equation~\ref{eqn:flux}) this means our size limit may need to be
  increased significantly.  However, knowledge of individual bright
  transmitters (from the type of analysis presented in this paper) means we can reduce the number of trials in the frequency axis
  substantially, and seeking patterns among adjacent time samples that
  fit plausible ephemerides will also help mitigate this effect.

If the challenges above can be addressed and a robust debris detection and tracking capability can be established with the MWA, it could become a very useful element of a suite of SSA facilities focussed on southern and eastern hemispheric coverage centered on Australia.

In November 2012 an announcement was made via an Australia-United States Ministerial Consultation (AUSMIN) Joint Communiqu\'{e} that a US Air Force C-band radar system for space debris tracking would be relocated to Western Australia\footnote{http://foreignminister.gov.au/releases/2012/bc\_mr\_121114.html}.  This facility will provide southern and eastern hemispheric coverage, allowing the tracking of space debris to a 10 cm size scale.  This system can produce of order 200 object determinations per day (multiple sets of range, range-rate, azimuth and elevation per object) and will be located near Exmouth in Western Australia, only approximately 700 km from the MWA site.

Additionally, the Australian Government has recently made investments into SSA capabilities, via the Australian Space Research Program (ASRP): RMIT Univerity's ``Platform Technologies for Space, Atmosphere and Climate''; and a project through EOS Space Systems Pty Ltd, ``Automated Laser Tracking of Space Debris''.  The EOS project will result in an upgrade to a laser tracking facility based at Mt Stromlo Observatory, near Canberra on Australia's east coast, capable of tracking sub-10 cm debris at distances of 1000 km\footnote{http://www.space.gov.au}.

In principle, the C-band radar system, a passive radar facility based on the MWA, and the laser tracking facility could provide a hierarchy of detection and tracking capabilities covering the southern and eastern hemispheres.  The C-band radar system could potentially undertake rapid but low positional accuracy detection of debris, with a hand-off to the MWA to achieve better angular resolution and rapid initial orbit determination (via long duration tracks).  The MWA could then hand off to the laser tracking facility for the most accurate orbit determination.  Such a diversity of techniques and instrumentation could form a new and highly complementary set of SSA capabilities in Australia.  A formal error analysis, describing the benefits of combining the data from such a diverse set of tracking assets will be the subject of work to be presented in a future paper.

Such a hierarchy of hand-off between facilities has implications for scheduling the MWA, whose primary mission is radio astronomy.  Currently the MWA is funded for science operations at a 25\% observational duty cycle (approximately 2200 hours of observation per year).  This duty cycle leaves approximately 6500 hours per year available to undertake SSA observations (assuming that funds to operate the MWA for SSA can be secured).  Even taking a conservative approach where only night time hours are used for observation, up to approximately 2200 hours could be available for SSA.  It would be trivial to schedule regular blocks of MWA time, per night or per week, to undertake SSA observations.  Given the rapid sky coverage of the MWA and the short observation times effective for SSA, the entire sky could be scanned over the course of approximately one hour each night, or using multiple shorter duration observations over the course of a night.  Since the MWA can be rapidly repointed on the sky, the possibility of interupting science observations for short timescale follow-up SSA observations, triggered from other facilities such as the C-band radar, could be considered.

A review of Australia's capabilities in SSA was presented by \citet{new11}, pointing out that better connections between the SSA community and latent expertise and capabilities in the Australian astronomical community could be leveraged for significant benefit.  In particular the authors suggested involving new radio astronomy facilities in SSA activities, given the advances in Western Australia connected with the SKA.  The development of a passive radar facility in the FM band using the MWA would be a significant step in this direction.

Immediate future steps are to repeat the ISS measurements with the full 128 tile MWA, as an improvement on the 32 tile array used for the observations reported here, as well as to target smaller known satellites as tests and to implement the improvements described above.

\acknowledgements
We acknowledge the Wajarri Yamatji people as the traditional owners of
the Observatory site.  Support for the MWA comes from the
U.S. National Science Foundation (grants AST CAREER-0847753,
AST-0457585, AST-0908884, AST-1008353, and PHY-0835713), the Australian Research Council (LIEF grants LE0775621 and LE0882938), the U.S. Air Force Office of Scientic Research (grant FA9550-0510247), the Centre for All-sky Astrophysics (an Australian Research Council Centre of Excellence funded by grant CE110001020), New Zealand Ministry of Economic Development (grant MED-E1799), an IBM Shared University Research Grant (via VUW \& Curtin), the Smithsonian Astrophysical Observatory, the MIT School of Science, the Raman Research Institute, the Australian National University, the Victoria University of Wellington, the Australian Federal government via the National Collaborative Research Infrastructure Strategy, Education Investment Fund and the Australia India Strategic Research Fund and Astronomy Australia Limited, under contract to Curtin University, the iVEC Petabyte Data Store, the Initiative in Innovative Computing and NVIDIA sponsored CUDA Center for Excellence at Harvard, and the International Centre for Radio Astronomy Research, a Joint Venture of Curtin University and The University of Western Australia, funded by the Western Australian State government.  The electromagnetic simulations were performed by Gary Bedrosian and Randy Ward of Remcom Inc.

{\it Facility:} \facility{MWA}.

\end{document}